\documentstyle[aps,prb,epsfig]{revtex}
\begin{document}
\twocolumn[
{\hsize\textwidth\columnwidth\hsize\csname@twocolumnfalse\endcsname
\draft
\title{Effect of unitary impurities 
in non-STM-types of tunneling in high-$T_c$ superconductors} 
\author{Jian-Xin Zhu and  C. S. Ting} 
\address{Texas Center for 
Superconductivity and Department of Physics, University of Houston, Houston, 
Texas 77204} 
\author{Chia-Ren Hu}
\address{Department of Physics, 
Texas A\& M University, College Station, Texas 77843}
\maketitle
\widetext
\begin{abstract}
Based on an extended Hubbard model, we present calculations of both the 
local ({\it i.e.,} single-site) and spatially-averaged differential 
tunneling conductance in $d$-wave superconductors containing nonmagnetic 
impurities in the unitary limit. Our results show that a random 
distribution of unitary impurities of any concentration can at 
most give rise to a finite zero-bias conductance (with no peak there) 
in spatially-averaged non-STM type of tunneling. This is in spite of the 
fact that local tunneling in the immediate vicinity of an isolated 
impurity does show a conductance peak at zero bias. We also find that to 
give rise to even a small zero-bias conductance peak in the 
spatially-averaged type of tunneling the impurities must form dimers, 
trimers, etc. along the $[110]$ directions. In addition, 
we find that the most-recently-observed novel pattern of the tunneling 
conductance around a single impurity by
 Pan {\em et al.} [Nature {\bf 403}, 746 (2000)] can be explained 
in terms of a realistic model of the tunneling configuration which 
gives rise to the experimental results reported there. 
The key feature in this model is the blocking effect of the BiO and SrO 
layers which exist between the tunneling tip 
and the CuO$_2$ layer being probed.

\end{abstract} 
}
]

\narrowtext
\def\neqto{\,\,/\llap=\,}
\section{Introduction}
Several years ago, one of us (CRH)~\cite{hu94} showed that the 
quasi-particle spectrum of a $d$-wave superconductor (DWSC) contains 
a special class of excitations --- called midgap states (MS's) in that 
reference --- which have essentially zero energy with respect to the 
Fermi energy, and are bound states with their wave functions localized 
at the vicinities of various kinds of defects in the system, such as 
a surface~\cite{hu94} and a grain~\cite{grainbdry-1,grainbdry-2,hu98} 
or twin~\cite{twinbdry} boundary. These MS's form an ``essentially 
dispersionless'' branch of elementary excitations, in the sense that 
their momenta along a flat surface or interface can essentially range 
from $-k_F$ to $k_F$ (Fermi momentum), and yet with almost no accompanied 
kinetic energy variation. The existence of surface and interface MS's 
appears to have already been confirmed by several types of 
experiments.~\cite{ZBCPexpt,lambda-anomaly,non-frauenhofer} An important 
question is whether a unitary impurity can also give rise to MS's. 
The MS's have a topological origin, in the sense that in the semi-classical
 WKBJ approximation, which makes these states truly midgap, 
their existence requires the satisfaction of one or more sign conditions 
only. More precisely, one can describe such a state in terms of one or 
more (as a linear combination) closed classical orbits, 
each of which must encounter two Andreev reflections by the pair potential
 at two different points of the Fermi surface where the pair potential
 ($\Delta(\vec k)$) has opposite signs. For each MS formed at a 
specular surface, only one such closed 
orbit is involved, so only one sign condition is required.~\cite{hu94}
 For a MS formed at a flat grain (or twin) boundary, modeled as a planar 
interface with 
transmissivity $0<t<1$ and different crystal orientations on its two sides,
 two closed classical orbits are involved, corresponding to the
possibility of transmission and reflection at the interface. Thus two
sign conditions must be satisfied simultaneously.~\cite{hu98}
 If a unitary impurity could be represented by a circular hole of a
 radius much larger than the Fermi wavelength, then quasi-classical
 argument can be expected to hold, and at least some non-vanishing number
 of midgap states should exist near its boundary. But when a unitary
 impurity is of atomic size, such a 
quasiclassical argument becomes dubious. 
In fact, if one thinks of scattering by an impurity as a linear
 combination of infinite number of
 classical orbits, 
corresponding to the possibility of scattering by all angles on the
Fermi surface, then it would seem that an infinite number of sign 
conditions would have to be satisfied in order for the impurity to 
induce some midgap states, corresponding to requiring sign change 
of the pair potential between any two points on the Fermi surface, 
which clearly is not satisfied. This argument suggests that midgap 
states can not form near an impurity, or at least they will not be 
exactly midgap even in the semiclasscal approximation. However, this
 argument does not distinguish between a unitary impurity and a 
non-unitary one. On the other hand, early treatments of a random
 distribution of unitary impurities in DWSC's, based on the 
self-consistent $t$-matrix approximation, have indicated a broad 
peak at zero energy in the density of states (DOS).~\cite{preosti94}
 (Even earlier similar studies on $p$-wave SC's also showed such 
peaks.~\cite{hrschfld88}) Perhaps Balatsky {\it et al.} are the 
first to mention within this context that a single {\em unitary} 
impurity in a $d$-wave SC will lead to two zero-energy bound states 
(per spin), with energies $\pm \epsilon_0$.~\cite{bltsky95} 
(Buchholtz and Zwicknagl made a similar statement much earlier for $p$-wave 
superconductors.~\cite{BZ91}) The $t$-matrix approximation employed in all 
of these works, which includes a semiclassical approximation, restores 
particle-hole symmetry, which is not a very good approximation in 
high-$T_c$ superconductors (HTSC's). 
Without this symmetry the energies of these ``MS's'' are most-likely
 not exactly zero. Unlike MS's at surfaces and interfaces, which 
involve a small number of points on the Fermi surface only, and 
therefore can easily avoid the gap-node directions, in the case 
of a state localized around an impurity, all points on the Fermi 
surface must be involved, including the nodal directions where the 
gap vanishes. Then for a state whose energy is not exactly zero, 
its wave function must be able to leak to infinity near the nodal 
directions.  That is, such a state can not be a genuine bound state, 
and can only be a resonant state. When there is a finite density of 
impurities, such wave functions should then possess a long-range 
interaction with each other via the ``leakages'', leading presumably 
to a broad ``impurity  band''. The self-consistent $t$-matrix approximation
 does not take this point into account properly, and is therefore not 
satisfactory. (However, see Joynt~\cite{joynt97} for an opposite view,
 except on the assumption of particle-hole symmetry and the validity 
of neglecting ``crossing diagrams'' when performing impurity averaging
 in two dimensions. But we think that neglecting the crossing diagrams 
is precisely why that approximation can not treat the 
impurity interaction properly.) A numerical approach has been introduced 
by Xiang and Wheatley~\cite{XW95} to avoid these shortcomings. For a single
 unitary impurity and a random distribution of unitary impurities it gave 
results in good agreement with the $t$-matrix approximation on the DOS. 
Numerical  solutions of a lattice BCS model with nearest neighbor attraction
 has also been employed by Onishi {\it et al.},~\cite{OOSM96} who showed 
that: (i) their exist two essentially-zero-energy states (per spin) localized
 around an impurity, (ii) their wave functions have long tails in the nodal 
directions, (iii) as a result such states localized around two impurities 
separated by a large distance in comparison with the coherence length 
$\xi_0$ can still interact with 
each other, leading to a broad impurity band. Thus the general picture 
outlined above appears to be confirmed. But Ref.~\onlinecite{XW95} 
probably also assumed particle-hole symmetry since it plotted the DOS 
for positive energy only, whereas in 
Ref.~\onlinecite{OOSM96} only $\mu=0$ is considered which has exact 
particle-hole symmetry. Numerical diagonalization method has also been 
applied to particle-hole non-symmetric models to study impurity effects, 
but without addressing the questions raised here. See, 
for example, Ref.~\onlinecite{franz96}. 
Rather, that work and several other works~\cite{lclztn} debated on 
whether there is localization in the impurity band --- a topic which 
is not our concern here. Instead, we wish to address the roles played by 
unitary impurities in single-particle tunneling. In particular, it has been
 noted that MS's formed at surfaces and interfaces of DWSC's can lead to an 
observable zero-bias conductance peak (ZBCP) in 
tunneling.~\cite{hu94,hu98,TK95,XMT96,ZT98,ZFT99} 
Indeed, several recent tunneling
 experiments performed with~\cite{ZBCP-STM} STM/S and other tunneling 
techniques \cite{other} on HTSC single crystals and epitaxial thin 
films have shown that in $ab$-plane tunneling a
very prominent ZBCP can be observed, especially on $\{110\}$ surfaces. 
The observed ZBCP exists continuously for long distances along the 
surfaces, and the observed tunneling characteristics can be 
quantitatively fitted by a generalized Blonder-Tinkham-Klapwijk 
theory~\cite{BTK} which includes the effects of the MS's formed 
on such surfaces of DWSC's. Assuming that this interpretation is
 correct, a question one can ask next is: Can unitary impurities 
be responsible for at least some of the observed ZBCP's?  
This question is meaningful since ZBCP's have been 
observed ubiquitously in all kinds of tunneling settings, and some 
of them may not possess surface and interface MS's. 
(See Ref.~\onlinecite{hu98} for a review.)  Theoretical results reviewed 
above, although not conclusive, seem to suggest that unitary impurities 
can also give rise to ZBCP's. But experimental evidence in non-STM/S 
types of tunneling seem to suggest the contrary. To see that this is 
the case, one must first exclude ZBCP's observed 
in $ab$-plane tunneling, and on tunneling performed 
on polycrystaline and ceramic samples, since in these cases contributions
 from the surface and interface MS's can most-likely dominate.  
(Whereas no MS's can form on a flat $\{100\}$ surface of a 
DWSC,~\cite{hu94} surface roughness can reverse this 
conclusion.~\cite{a-axis} Thus we exclude $a$-axis and 
$\{n0m\}$-directional tunneling as well, if $n \neqto 0$.) 
One is left with $c$-axis tunneling on single crystals and 
epitaxial thin films only. (Nominal $c$-axis point-contact measurements
 may actually be seeing some $ab$-plane tunneling, since the tunneling 
tips have been pushed into the HTSC samples in these measurements. 
So they should also be excluded.) Even in the surest $c$-axis 
tunneling cases one must
 still distinguish between STM/S-type tunneling, which can 
explore the tunneling characteristics in the close vicinity of a  
single isolated impurity, and the other non-localized tunneling 
techniques, which see a spatially averaged tunneling characteristics. 
For the later type, the spectral weight of the impurity contribution  
must not be too low to be observed, so the impurity concentration must
not be too low. It is this kind of tunneling which we are interested 
in here, since we suspect that once the spectral weight is sufficiently 
large, the interaction between a random distribution of impurities will 
also be so large that it still can not give a ZBCP, 
but only a finite conductance 
at zero bias, $G(0)$, as a local minimum or even an extra dip.  
Searching the literature, we find three more recently published 
papers reporting the STM/S results on the observation of a ZBCP-like
 feature in the vicinity of an 
impurity.~\cite{Yazdani99,Hudson99,Pan99} 
For non-localized tunneling, we find at most a few cases which can weakly 
suggest that the small ZBCP's observed in them might originate from 
impurities.~\cite{ZBCP-imp} (Even in these cases, it is not clear 
whether MS's could have formed at some exposed CuO$_2$ edges at the 
interface with the insulating barrier.  The epitaxial films might also 
have grain boundaries which could host MS's.)  Most $c$-axis tunneling 
data which exhibit a clear gap feature show a minimum at zero bias, with
 some showing essentially simple $d$-wave behavior, with very small 
$G(0)$,~\cite{zerog0} and some showing nearly $d$-wave behavior but 
with a finite $G(0)$.~\cite{finiteg0} Still 
others show features on both sides of zero bias, giving the impression
 that there is an extra dip at zero bias.~\cite{exdip} Very few 
publications seem to have systematically studied the impurity 
effects in $c$-axis tunneling. We find one:  
Hancotte {\it et al.}~\cite{hancot97} showed that Zn substitution
 ($\le 1\%$) in the CuO$_2$ planes of BSCCO (2212) caused $G(0)$ 
to markedly increase, accompanied by a reduced gap. Here the spectral
 weight of impurity effects is clearly large enough, yet not even a 
trace of a ZBCP was observed. 

More recently, the quasiparticle properties around a single impurity 
have been investigated in more detail in a two dimensional $t$-$J$ 
model,~\cite{Tsuchiura99,Zhu99} with a focus on whether the ZBCP 
observed with STM/S near an impurity is split or not.  The purpose 
of this paper is, on the other hand, to attempt to answer 
the following precise question: Assuming that HTSC's are DWSC's, 
and isolated unitary impurities do possess near-zero energy resonant
 states which can be observed as a ZBCP-like feature by STM/S in the 
close vicinity of such an impurity, can any concentration of a random
 distribution of unitary impurities be able to 
give rise to an observable ZBCP (of any width) in  non-STM/S types
 of tunneling, or there must be some spatial correlation in the
 impurity distribution before a ZBCP can appear in such types of 
tunneling? For this purpose we have performed an 
extensive numerical study. We have introduced the supercell technique
 so that the finite-size effects from the exact diagonalization can 
be overcome, and the desired energy resolution can be obtained. 
This technique has the ability to treat well the impurity of 
atomic size. Moreover, the band structure effects can be incorporated
 in a natural way. The results show the conductance behavior is 
sensitive to the position of the chemical potential within the 
band and to the impurity configuration at the atomic scale. Our
 results have indeed confirmed our suspicion that for a simple random
 distribution of unitary impurities, either their spectral weight is 
too low for their effects to be observable in non-STM types of tunneling,
 or their interaction is so strong that only
 a finite $G(0)$ is obtained as a local minimum rather than a peak,
 because the impurity band has spread wide, with the center of its 
contribution to the density-of-states function lower than its two sides.
 In addition, we find that if only enough number 
of 
the unitary impurities form nearest neighbors along the $[11]$ 
directions in a CuO$_2$ plane, henceforth called 
``$[11]$-directional dimers'', then a weak ZBCP can appear in 
non-STM types of tunneling.  
(We also find that if enough such impurities form $[11]$-directional  
{\it trimers}, then the ZBCP can be even taller and narrower, but the 
chance of forming such alignments in an actual sample is probably very 
low. On the other hand, we think that dimers can probably form with not
 very low probability.) Furthermore, we 
have also shown that the most recently observed pattern (by STM/S) of 
the local tunneling 
conductance around a single impurity~\cite{Pan99} can be explained by 
taking into account that the STM/S tip in that experiment is separated
 from the CuO$_2$ plane under probe by a BiO layer and a SrO layer. 
Therefore, the tunneling tip can not communicate with the atom 
directly below it in the CuO$_2$ plane, due to the 
blocking effect of the atoms directly above it in the BiO and SrO layers.
 Rather, we think that the measured ``local tunneling conductance'' by 
the tunneling tip on a Cu site in the CuO$_2$ plane is actually that 
averaged over a local region around that site, 
{\it excluding the contribution from the atom at that site, 
because of the blocking effect just described}. Using this 
very reasonable postulate, we find that we can at least 
semi-quantitatively understand the pattern observed in 
Ref.~\onlinecite{Pan99}, including why it peaks at the impurity site,
 and vanishes at its four nearest neighbor Cu sites, etc.
    
\section{Theoretical Method}
To model decoupled copper-oxygen layers in HTSC's, we consider 
the single-band extended Hubbard model defined on a two-dimensional 
square lattice (lattice constant $a$) with nearest-neighbor hopping,
 and on-site repulsive and nearest-neighbor attractive 
interactions.~\cite{XW95} For our purpose, we introduce supercells 
each with size $N_{x}a\times N_{y}a$. We then define the supercell 
Bloch states labeled by a wave vector ${\bf k}$ and a site index 
${\bf i}$ within the supercell. In the mean field 
theory, the task becomes to exactly diagonalize the  
Bogoliubov-de Gennes (BdG) equations:~\cite{deG66}
\begin{equation}
\label{EQ:BdG}
\sum_{\bf j} \left( \begin{array}{cc}H_{\bf ij}({\bf k}) 
& \Delta_{\bf ij}({\bf k}) \\
\Delta_{\bf ij}^{\dagger}({\bf k}) & -H_{\bf ij}({\bf k})\end{array} \right) 
\left( \begin{array}{c} u_{\bf j}^{n,{\bf k}} \\ 
v_{\bf j}^{n,{\bf k}} \end{array} \right)
= E_{n,{\bf k}} \left( \begin{array}{c}
u_{\bf i}^{n,{\bf k}} \\ v_{\bf i}^{n,{\bf k}}\end{array} \right)\;.
\end{equation}
Here $u_{\bf i}^{n,{\bf k}}$ and $v_{\bf i}^{n,{\bf k}}$ are the  
Bogoliubov amplitudes corresponding to the eigenvalue $E_{n,{\bf k}}$.
\begin{equation}
H_{\bf ij}({\bf k})=-te^{i{\bf k}\cdot\mbox{\boldmath{$\delta$}}a} 
\delta_{{\bf i} + \mbox{\boldmath{$\delta$}},{\bf j}}
 + (U_{\bf i}-\mu)\delta_{\bf ij}\;,
\end{equation}
\begin{equation}
\Delta_{\bf ij}=\Delta_{0}({\bf i})\delta_{\bf ij} 
+ \Delta_{\mbox{\boldmath{$\delta$}}}({\bf i})
e^{i{\bf k}\cdot \mbox{\boldmath{$\delta$}} a}
\delta_{{\bf i} + \mbox{\boldmath{$\delta$}},{\bf j}}\;,
\end{equation}  
where $t$ is the hopping integral, $\mu$ is the chemical potential,
$\mbox{\boldmath{$\delta$}}=\pm \hat{\bf x}, \pm \hat{\bf y}$ 
are the unit vectors along the crystalline $a$ and $b$ axes,
and $k_{x,y}=2\pi n_{x,y}/M_{x,y}N_{x,y}a$ with
$n_{x,y}=0,1,2,\dots,M_{x,y}-1$. $M_{x,y}N_{x,y}a$ is the linear 
dimension of the whole system, which is assumed to be made of $M = 
M_x\times M_y$ super-cells. The single-site nonmagnetic impurity 
scattering is represented by $U_{\bf i} = U_{0}\sum_{{\bf i}^{\prime}
\in I}\delta_{{\bf i}^{\prime},{\bf i}}$ with the summation over the
 set of impurity sites. The self-consistent pair potentials are in 
turn expressed in terms of the wavefunctions $(u_{\bf i},v_{\bf i})$:
\begin{equation}
\label{EQ:Gap0}
\Delta_{0}({\bf i})=\frac{g_{0}}{M}\sum_{n,{\bf k}} u_{\bf i}^{n,{\bf
k}}(v_{\bf i}^{n,{\bf k}})^* \tanh (E_{n,{\bf k}}/2k_{B}T)\;,
\end{equation} 
\begin{eqnarray}
\Delta_{\mbox{\boldmath{$\delta$}}}({\bf i}) &=& 
\frac{g_{1}}{2M}\sum_{n,{\bf k}}[u_{\bf i}^{n,{\bf k}}(v_{{\bf i} + 
\mbox{\boldmath{$\delta$}}}^{n,{\bf k}})^*e^{-i{\bf k}\cdot
\mbox{\boldmath{$\delta$}}a} + u_{{\bf i} + 
\mbox{\boldmath{$\delta$}}}^{n,{\bf k}}(v_{\bf i}^{n,{\bf k}})^*
e^{i{\bf k}\cdot\mbox{\boldmath{$\delta$}}a}]\nonumber \\
&&\times \tanh (E_{n,{\bf k}}/2k_{B}T)\;,
\label{EQ:Gap1}
\end{eqnarray}  
where $k_{B}$ is the Boltzmann constant and $T$ 
is the absolute temperature. In the case of repulsive 
on-site ($g_0<0$) and nearest-neighbor attractive ($g_1>0$)
 interactions, the $d$-wave pairing state is favored, the 
amplitude of which is defined as:
$\Delta_{d}({\bf i})=[\Delta_{\hat{\bf x}_a}({\bf i}) + 
\Delta_{-\hat{\bf x}_a}({\bf i}) - \Delta_{\hat{\bf x}_b}({\bf i}) -
\Delta_{-\hat{\bf x}_b}({\bf i})]/4$.

We define the local differential-tunneling-conductance (DTC) 
by~\cite{Tinkham75} 
\begin{eqnarray}
G_{\bf i}(E) = -\frac{2}{M}\sum_{n,{\bf k}}
[\vert u_{\bf i}^{n,{\bf k}}\vert^{2}f^{\prime}(E_{n,{\bf k}}-E)\nonumber \\ 
+ \vert v_{\bf i}^{n,{\bf k}}\vert^{2}f^{\prime}(E_{n,{\bf k}}+E)]\;,
\end{eqnarray}
where the prefactor $2$ comes from the two-fold spin degeneracy and 
$f^{\prime}(E)$ is the derivative of the Fermi distribution function 
$f(E)=[\exp(E/k_{B}T)+1]^{-1}$. $G_{\bf i}(E)$ can theoretially be 
measured by STM experiments. (However, see later for a possible 
complication when the probed layer and the STM/S tip are separated by
other atomic layers.) The spatially-averaged DTC is defined by 
$G(E)\equiv \sum_{\bf i} G_{\bf i}(E)/N_x N_y$, which can essentially
 be measured in many non-STM/S types of tunneling experiments (planar,
 ramp, etc.). (But squeezable junctions might measure something in between,
 depending on the pressure applied.)

\section{Numerical Results}
In the numerical calculation, we take the supercell size $N_x=N_y=35$, 
the number of supercells $M=6 \times 6$, and the temperature 
$k_{B}T=0.02t$.  The lattice sites within one supercell are indexed as 
$(i_x, i_y)$ with $i_x$ and $i_y$ each ranging from 1 to 35. 
In addition, we use the bulk value of the order parameter as 
an input to diagonalize Eq.~(\ref{EQ:BdG}) and the deformation
 of the order parameter near the impurity is ignored. (This 
approximation should be quite acceptable for the questions we wish to 
answer here.) In the bulk system, the $d$-wave order parameter
 has the form $ \Delta_{\bf k} = 2\Delta_{d} (\cos k_{x}a -\cos
 k_{y}a)$, where $k_{x,y}$ are the $x$- and $y$-components of the
 wave vector defined on the whole system. Table~\ref{TABLE:I} lists
 the resulting $d$-wave pair potential $\Delta_{d}$ and coherence 
length $\xi_0 = \hbar v_{F}/\pi  \Delta_{\mbox{\small max}}$ for 
the chosen values of $g_1$ and $\mu$, where $v_{F}$ is the Fermi 
velocity and $\Delta_{\mbox{\small max}} = 4\Delta_{d}$ is 
the maximum energy gap. The choice of the maximum energy gap
 is consistent with the bulk gap structure exhibited in the 
bulk DOS. In the calculation of $\Delta_d$, an arbitrary 
negative value of $g_0$ can be taken. 

\subsection{The cases of a single impurity and a small cluster of impurities}
The quasiparticle property near a single impurity in a DWSC is 
complicated, and whether the energy of quasiparticle resonant 
states is exactly zero (relative to the Fermi energy) or whether the zero (or 
near-zero) energy states are split is very sensitive to the band 
structure and the impurity strength. To give a clear answer or clue to 
this question, we first consider the case of weak or moderately strong 
impurity. Figure~\ref{FIG:LDOS-0} plots the local DTC as a function of 
bias directly on the single-site impurity (a) and one lattice constant 
away (b). As shown in Fig.~\ref{FIG:LDOS-0}, when the impurity scattering
 is weak ($U_0=2.5t$), the local DTC on the impurity site has a peak below
 the Fermi energy, which is consistent with the earlier study within the 
continuum theory;~\cite{bltsky95} 
while that at the site nearest-neighboring to the impurity has a 
double peak structure, one above and the other below the Fermi energy.
 When the impurity scattering becomes stronger ($U_0=10t$), the peak 
on the local DTC on the impurity site is pushed 
toward the Fermi energy with the amplitude strongly suppressed;
 nevertheless the double peaks on the local DTC at the nearest-neighbor
 site converge to each other and the intensity is enhanced. Since $\mu=0$
 in this calculation, the band is globally 
particle-hole symmetric. Therefore the non-zero energy resonant states
 shown above originates from the local particle-hole symmetry breaking.
 Differently, when the impurity scattering goes to the unitary limit, 
whether there exists the particle-hole symmetry 
depends on solely on the position of chemical potential within the band.  

In all the following discussions, the single-site impurity potential 
strength is taken to be $U_0=100t$, so it is practically in the unitary
 limit. In Fig.~\ref{FIG:LDOS-1} the local DTC is plotted as a function
 of bias near one single-site impurity located 
at site $(18,18)$ for various values of $\mu$ but 
with $g_1 = 2t$ fixed. The bias is normalized to 
$\Delta_{\mbox{\small max}}$. The tunneling point is one 
lattice constant away from the impurity along the $(10)$ 
direction, i.e., at (19, 18). As is shown, when $\mu = 0$ where the 
particle-hole symmetry holds, a sharp zero-bias conductance peak is 
exhibited. When $\mu$ deviates from zero, the line shape of the 
local DTC becomes asymmetric with respect to the zero energy position. 
The peak near zero energy is seen to show an incomplete 
splitting and the height of this peak is decreased. The extent of splitting 
increases with the deviation of $\mu$ from zero. This 
result is quite different from the case of a $\{110\}$-oriented 
surface, where an un-split zero-energy peak in the local DTC shows
 up regardless of the position of the chemical potential. The 
similarity between the result obtained here by solving the extended 
Hubbard model and the corresponding result we obtained earlier by 
solving the $t$-$J$ model~\cite{Zhu99} indicates that our conclusion
 about the splitting of the ZBCP is, at least in the mean 
field level, model independent. Next, we show that when more 
than one unitary impurities are present in the SC, the local 
DTC heavily depends on the impurity configuration. In 
Fig.~\ref{FIG:LDOS-2}, we plot the local DTC 
in a DWSC with two and three impurities forming a $[11]$-directional 
nearest-neighbor dimer and trimer, respectively, for $g_1=2t$ 
and $\mu=-t$. Without loss of generality, the impurity positions are 
placed at sites (17,19) and (18,18) for the dimer case, and at
 sites (17,19), (18,18) and (19,17) for the trimer case. The 
measured position is still at site (19,18). As can be seen clearly,
 the near-zero-energy peaks in the local DTC for the single 
impurity case are strongly pulled toward zero energy in the dimer
 case. In the trimer case, the zero-energy peak is even more 
pronounced. To check this point more seriously, we have also 
calculated the lowest positive eigenvalues for the system composed of 
one supercell with one single impurity, and two impurities with different
 relative positions for $g_1=2t$ and  $\mu=-t$. We find that a single 
impurity leads to two near-zero-energy states per spin, one just above
 and one just below the Fermi energy. The 
absolute energy corresponding to these two eigenstates is roughly
 $0.1t$.  (It is not zero because there is no particle-hole symmetry
 in this model study.) But for the two impurities positioned as nearest
 neighbors of each other along the $[11]$ direction, 
the lowest positive eigen-energy is only about $0.03t$. If two impurities
 are positioned as nearest neighbor of each other along the $[10]$ 
direction, the lowest positive eigenvalue is roughly $0.09t$, which
 is very close to the value for isolated 
impurities. In Fig.~\ref{FIG:DISTANCE}, we have plotted the variation
 of the lowest positive eigenvalue with the distance between two 
imputities aligned along the $[11]$ direction. When two impurities
 are far apart, the lowest positive eigenvalue oscillates 
between $\sim 0.105t$ and $\sim 0.079t$, around the value $\sim 0.1t$
 for isolated impurities, as the distance between the two impurities
 are increased. This oscillation is a clear indication of the long 
range interaction  
between two impurities that we have already discussed. (The long tails
 of the wave function of a bound state around a unitary impurity in the 
nodal directions are oscillating in Fermi wavelength, so the sign of  
the interaction between two impurities can change with distance.) When
 two impurities form a $[11]$-directional dimer, they combine to play 
the role of a very short $\{11\}$ edge. That is, they drastically 
enhanced the probability for specular reflection 
by this edge, so the nearest-to-zero eigen-energies become very close to
 zero energy. This analysis demonstrates that the closest-to-zero 
eigen-energies due to the presence of impurities are very sensitive
 to the short-range correlations in the impurity 
configuration.  Furthermore, we plot in 
Fig.~\ref{FIG:NUMBER} the number of near-zero eigen-energy states versus 
the number of impurities aligned consecutively along the $[11]$ 
direction.  In this calculation, each state with its eigenvalue smaller 
than $0.04t$ is counted as a near-zero-energy state. As the result shows,
 the number of near-zero-energy states increases almost 
linearly with the length of a $[11]$-directional impurity line. We also 
find that, as the number of impurities is increased, there are more and 
more states with their eigen-energy approaching zero, which confirms from
 an alternative aspect that the 
appearance of the ZBCP in the $\{110\}$-oriented DWSC tunnel
 junctions is independent of the position of $\mu$ in the band.  
 
Most recently, Pan {\em et al.}~\cite{Pan99} have performed an STM
 study on the effect of a single zinc (Zn) impurity atom on the 
quasiparticle local density of states (LDOS) in BSCCO. Besides 
revealing the predicted highly localized four-fold 
quasiparticle cloud around the impurity, the imaging 
also exhibited a novel distribution of the near-zero-resonant-energy LDOS
 near the impurity. In contrast to the existing theories, which give a 
vanishing LDOS directly at the site of the unitary impurity, and 
vanishing or low values on all atomic sites along the $[11]$ 
directions and at the $(20)$ and $(02)$ sites, it was observed
 that the LDOS at the resonance energy has the strongest intensity
 directly on the Zn site, scattering from which is believed to 
be in the unitary limit. In addition, the LDOS at the resonance 
energy has nearly local minima at the four Cu atoms nearest-neighbor to the
 Zn atom, and has local maxima at the second- and third-nearest-neighbor
 Cu atoms. The intensity of the LDOS on the 
second-nearest-neighbor Cu atoms is larger 
than that on the third-nearest-neighbor Cu atoms.
 Although this unexpected phenomenon might indicate
 strong correlation effects, we would rather try to
 explain it from an alternative point of view. 
Superconductivity in 
HTSC's is believed to originate in the CuO$_2$ plane, 
and the STM/S tip at low bias is known to be probing such a plane closest
 to the surface. But the STM tip is also known to be separated from this
 plane by a BiO layer and a SrO layer. The BiO layer is 
believed to be semiconducting and the SrO layer, insulating.
 So low-energy electrons can go through them. However, the hard cores of
 the atoms surely can block the tunneling current from directly going
 through them. We therefore postulate that tunneling 
occurs from the tip to the atoms within a small circular area in the CuO$_2$
 plane directly below the tip, except those atoms blocked in the above sense.
 In addition, we postulate that the linear dimension of the small circular 
area is only a little larger than the lattice constant $a$, because 
the negative-exponential dependence of the tunneling current on distance
 implies that we need only include the closest set of atoms that {\it can}
 contribute to the tunneling current. (That is, they are not blocked 
by the atoms in the BiO and SrO layers closer to the surface than the
 CuO$_2$ layer being probed.) Then the tip above a given Zn or Cu site
 in the CuO$_2$ plane sees only its four nearest-neighbor sites in that
 plane. Therefore, we propose that the 
actual measured LDOS at a Zn or Cu site is essentially the sum of
 the contributions from its four nearest (Cu or Zn)-neighbors. 
The measured local DTC should then be related to the calculated 
local DTC by the relation: 
\begin{equation}
\label{EQ:transf}
G_{\bf i}^{\mbox{\small expt.}}(V)
= \sum_{\mbox{\boldmath{$\delta$}}} G_{{\bf i} + 
\mbox{\boldmath{$\delta$}}}^{\mbox{\small calc.}}(V).
\end{equation} 
In Fig.~\ref{FIG:DISTRIBUTION}, part (a), the spatial distribution of 
the calculated bare [{\it i.e.}, before the transformation given in 
 Eq.~(\ref{EQ:transf})] DTC at zero bias is displaced in a three-dimensional
 plot. It includes the effects of all 
four near-zero-energy resonant states (two per spin) localized around a 
unitary impurity.~\cite{remark} The temperature is assumed to be at 
$k_BT = 0.02t$ (or $T \simeq 0.1 T_c$) in the calculation. The parameter
 values $g_1=1.5t$ and $\mu=-0.4t$ are chosen 
to obtain a single near-zero-bias conductance peak. These values are not
 yet optimized, since at the present time we only wish to establish the 
essential correctness of our idea.
In Fig.~\ref{FIG:DISTRIBUTION}, part (b), a planar "bubble plot" of the 
same data is given, where the size of each black dot is directly 
proportional to the calculated bare-DTC at that lattice site. 
(Our calculation, being tight-binding in nature, gives 
DTC values only at the lattice sites, unlike the observed data,
 which gives a continuous variation of the DTC between the atomic sites.)
 In Fig.~\ref{FIG:DISTRIBUTION}, parts (c) and (d), similar plots are given
 for the calculated transformed-DTC based on 
Eq.~(\ref{EQ:transf}). It is clear from comparing these plots with Fig. 3(b)
 and (c) of Ref.~\cite{Pan99} that the calculated transformed-DTC-distribution
 agrees quite well, at least qualitatively, with the measured 
DTC-distribution at the resonant energy 
in that reference.

As a crude quantitative comparison between our prediction 
based on Eq.~(\ref{EQ:transf}) and the measure data of 
Pan {\it et al.}, we have listed in Table~\ref{TABLE:II}
 the normalized measured values by Pan {\it et al.} 
[based on Fig.4 (a) in Ref.~\cite{Pan99}], 
our normalized calculated bare values, and our
 normalized calculated transformed values [by using 
Eq.~(\ref{EQ:transf})], of the local DTC at various lattice
 sites near a Zn impurity [which is defined to be  the (00)
 site], up to the third 
nearest-neighbor [{\it i.e.}, the (20) and (02)] sites. Within each
 row of data, the normalization is such that the
largest value becomes unity. (For the first and third rows of data,
 this occurs at the (00) site, but for the second row of data, this
occurs at the (10) and (01) sites, because the calculated bare value
of the local DTC at the resonant energy vanishes at the (00) site.)
We admit that this comparison is only a very crude one, since our 
tight-binding result for the DTC distribution, which exists at the
 Zn and Cu sites only, and not continuously in between them, should,
 strictly speaking, be compared with some integrated 
result of the measured local DTC. 
But in the first row of Table~\ref{TABLE:II} we have only listed the 
measured local DTC values right at the Zn and near-by Cu sites, with 
no integration performed. In fact, we strongly believe that the agreement
 will be 
much better if we do perform properly such an integration of the
 measured data before comparison with our prediction, as may be
 seen by comparing our prediction given in Fig.~\ref{FIG:DISTRIBUTION}, 
part (d), and the Fig. 3(b) of Ref.~\cite{Pan99}, or the three-dimensional
 figure on the cover page of the March 2000 issue of Physics
 Today~\cite{PhysToday}. This is particularly so with regard
 to the relative heights of the local DTC measured at the (10)
 or (01) sites, and those at the (20) and (02) sites. The
 former is larger than the latter in the data given in the
 first row of Table~\ref{TABLE:II}, and in Fig. 4(a) of 
Ref.~\cite{Pan99}, where the data in the first row of 
table~\ref{TABLE:II} came from. 
However, since at the moment
we have not figured out the proper way to do this integration, 
we shall leave that to a future publication. (An even better theory
 should generate 
a continuous local DTC in a plane, to be directly compared with the
 measured data, with no integration needed. This possibility will 
also be looked into later.) In spite of the fact that the comparison
 presented in Table II is only a very crude one, we 
believe it still strongly suggests that our idea is 
essentially correct, although the model clearly can and should be
 improved, in order to give a local DTC distribution which exists
 {\it continuously} in a two-dimensional plane, as has been observed. 
 
\subsection{The case of randomly distributed impurities}
In Fig.~\ref{FIG:LDOS-AV}, we have plotted the {\it spatially averaged}
 DTC for different concentrations of impurities contained in the DWSC. 
The parameters chosen are: $g_1=2t$ and $\mu=-t$. When the super-cell 
contains only one impurity, because the 
spectral weight from the impurities is too small for this density
 ($0.08\%$) of impurities, only a very small spatially-averaged 
zero-bias DTC, $G(0)$, appears, without a slight trace of a ZBCP.
 At this low concentration of impurities, $G(V)$ practically 
reflects the bulk $d$-wave DOS. The conductance peak outside 
the positive maximum energy gap stems from the van Hove singularity.
 To increase the spectral weight of the impurity contribution to $G(V)$,
 one needs to increase the density of impurities, in 
presumably a random distribution. But as shown by the dashed
 line in Fig.~\ref{FIG:LDOS-AV}, in which the randomly-distributed
 impurity density has been increased to $1.28\%$, one still does 
not obtain an observable ZBCP in spatially-averaged $G(V)$, but only
 a finite spatially-averaged $G(0)$ as a local minimum. This we think
 is because the wave functions of the near-zero-energy resonant states
 around the impurities have tails along the near-nodal directions, the
 interaction between such states at different 
impurity positions (especially between those impurities
 not very far away from each other) is so large that the
 contributions from these states to the DTC has spread into
 a wide band with actually lower value at the band center 
(i.e., zero bias) than at 
the band edges. This is consistent with the fact that in most
 $c$-axis non-STM-types of tunneling experiments, only a finite $G(0)$ is 
observed, without showing a peak there of any shape and width.  

Compared with the single impurity case plotted by the solid line,
 it is clear that $G(0)$ is enhanced at this much higher density
 of randomly-distributed impurities, showing that the spectral 
weight of the impurity contribution at this density of 
impurities is clearly no longer negligible. Yet no trace
 of a ZBCP is obtained. On the other hand, we find that at the
 same concentration of impurities, if a good portion of these 
impurities form $[11]$-directional dimers, (with separation 
$\sqrt{2}a$ between the
 two impurities in each such dimer,) but still with random orientations,
 a small observable ZBCP is exhibited in $G(V)$. [See the short-dashed 
line in Fig.~\ref{FIG:LDOS-AV}], in which  $62.5\%$ of the impurities 
form such dimers. We have also calculated 
the wavefunction near the two impurities forming such a dimer and found
 that the wavefunction  
amplitude is very small along the alignment direction of the two  
impurities, but has an oscillatory behavior perpendicular to the 
alignment direction. This highly anisotropic behavior of the 
wavefunction may have drastically suppressed the interaction 
between the impurities. The height of the ZBCP depends on the total 
number of impurities forming such dimers. We also find that if some 
impurities form $[11]$-directional {\it trimers}, the amplitude of 
the ZBCP can be enhanced even further. But we deem the probability 
for three impurities to form such a trimer in an 
actual system is very small, but the assumption that some
 impurities can form $[11]$-directional dimers should be
 reasonable. Thus we propose that this is how impurities
 can possibly contribute to some of the observed ZBCP's,
 and why often ZBCP's are not 
observed in $c$-axis non-localized tunneling, even in
 samples which have been deliberately introduced some
 substantial amount of substitutional non-magnetic zinc
 (Zn) impurities.~\cite{hancot97}

\section{Summary}
In summary, we have made detailed calculations of both the local and the 
spatially-averaged differential tunneling conductance in DWSC's  
containing nonmagnetic impurities in the unitary limit. Our results show 
the following: 
(1) Previously we have shown~\cite{Zhu99} that the local conductance 
behavior near zero energy at the sites near a unitary impurity is 
sensitive to how well the particle-hole symmetry is satisfied, 
and how large is the error of using a semiclassical WKBJ approximation
 to treat the problem. Here we have demonstrated  
that conclusion is model independent. (2) We also find that the conductance
 behavior is very sensitive to the impurity configuration. For the 
single-impurity 
case, a recently obtained LDOS imaging at the resonance energy has been
 explained in terms of a model where the blocking effect of the atoms in the
BiO and SrO layers are taken into account, so that the tunneling tip does 
not probe the local density of states of the Cu or Zn site directly below 
it, but rather essentially the sum of those of its four nearest neighbor 
sites. Furthermore, our study allows us to conclude that unitary impurities
 can contribute to an observable ZBCP in non-localized tunneling 
if and only if a substantial number of impurities form $[11]$-directional
 dimers, (and trimers, etc.). On the other hand a simple random distribution
 of unitary impurities either has too low a spectral weight to contribute 
observably to non-localized 
tunneling, or their interaction is already so strong that they can only
 produce a finite value at zero bias in non-localized tunneling conductance
 without giving a peak there of any shape or width. An ultimate test of this
 scenario would require: (i) the 
observation of a ZBCP in STM/S tunneling in the close vicinity of
 a unitary impurity, (which is already achieved recently); (ii) 
the non-observation of a ZBCP in non-STM/S types of $c$-axis tunneling
 on a single-crystal sample with any concentration of the 
same type of impurities which are confirmed to not have formed
 $[11]$-directional dimers, (or trimers, etc.); and (iii) the 
appearance of a small ZBCP in the same types of tunneling experiments
 when the impurity distribution is established to contain such 
dimers, (or trimers, etc.). Such an ultimate test may be asking too much
 from the experimentalists, but the fact that this scenario is consistent
 with many diverse tunneling observations give us much confidence on it 
being at least close to the truth.
 
\acknowledgments
We thank Dr. J. C. Seamus Davis and Ms. Kristine Lang for providing 
us quantitative information about the measurement they published in their 
recent Nature article.~\cite{Pan99} This work was supported by the Texas 
Center for Superconductivity at 
the University of Houston, the Robert A. Welch Foundation, the Grant 
under the No. NSF-INT-9724809, and by the
Texas Higher Education Coordinating Board under grant No. 1997-010366-029.

\begin{table}
\caption{The values of $d$-wave order parameter and coherence length for
several sets of parameter values. The temperature is at $T=0.02t$. }
\begin{tabular}{cccc}
$\mu/t$ & $g_{1}/t$ & $\Delta_d/t$ & $\xi_0/a$ \\
\tableline
0 & 2  & 0.241 & 1.33 \\
-0.4 & 2 & 0.229 & 1.32   \\ 
-0.4 & 1.5 & 0.139 & 2.17 \\
-1 & 2 & 0.164 & 1.68  
\end{tabular}
\label{TABLE:I}
\end{table}

\begin{table}
\caption{Comparison between the measured STM/S local differential tunneling
 conductance at the near-zero-bias resonant energy by Pan 
{\it et al.}~\cite{Pan99} at the Zn impurity site (00), 
the nearest neighbor sites (10) and (01), the next nearest
 neighbor site (11), and the third nearest neighbor sites (20) and (02),
 (the first row), and the calculated values for the local differential 
tunneling conductance at zero bias at the same sites before the transformation
 (according to Eq.~(\ref{EQ:transf})) 
(the second row), and after the transformation (the third row). The data 
in each row is normalized so that the largest value is unity, 
which occurs at the (00) site in the first and the third rows, but
 at the (10) and (01) sites in the second row.
}
\begin{tabular}{cccc}
$(00)$ & $(10)$ \& $(01)$ & $(11)$ & $(20)$ \& $(02)$ \\
\tableline
1.00 & 0.18 & 0.29 & 0.13 \\
0.000 & 1.000 & 0.078 & 0.113 \\
1.000 & 0.068 & 0.593 & 0.384 \\
\end{tabular}
\label{TABLE:II}
\end{table}

\begin{figure}
\caption{ 
The local differential tunneling conductance as a function of a bias at 
the site directly on the site (18,18) (a) and (19,18) (b) in a DWSC with 
each supercell containing one single impurity located at site (18,18). 
The impurity strength $U_0=2.5t$  (solid line) and $10t$ (dashed line). 
The other parameters $\mu=0$ and $g_1=2t$.
}
\label{FIG:LDOS-0}
\end{figure}

\begin{figure}
\caption{
The local differential tunneling conductance as a function of bias in a
DWSC with each supercell containing  one single 
impurity located at site (18,18) for $\mu = 0$ (solid line), $-0.4t$ 
(dashed line), and $-t$ (short-dashed line). The  parameter value of 
$g_1$ is taken to be $2t$. The measure point is at (19,18).
} 
\label{FIG:LDOS-1}
\end{figure}

\begin{figure}
\caption{
The local differential tunneling conductance as a function of bias in a 
$d$-wave superconductor with each supercell containing 
one single impurity 
(solid line) at site (18,18), and two (dashed line) at sites (17,19) and 
(18,18), and three impurities (short-dashed line) at sites (17,19), (18,18) 
and (19,17) aligned consecutively along the $\{110\}$ direction.  The 
parameters $g_1=2t$ and $\mu=-t$.  The measure point is at site (19,18).   } 
\label{FIG:LDOS-2}
\end{figure}

\begin{figure}
\caption{
The lowest positive eigenvalue as a function of the distance between two 
impurities aligned along the $\{110\}$ direction in the CuO$_2$ plane. The 
parameters $g_1=2t$ and $\mu=-t$.
}
\label{FIG:DISTANCE}
\end{figure}

\begin{figure}
\caption{
The number of near-zero energy states versus the number of impurities 
aligned consecutively along the $\{110\}$ direction in the CuO$_2$. 
The parameters $g_1=2t$ and $\mu=-t$.
}
\label{FIG:NUMBER}
\end{figure}

\begin{figure}
\caption{
Spatial distribution of the differential tunneling conductance at zero 
bias: The three-dimensional display in the whole supercell before 
summation over four nearest-neighboring Cu sites (a) and the 
corresponding two-dimensional bubble view in a small region  near the 
impurity (b); the three-dimensional display after summation (c) and the 
corresponding two-dimensional bubble view (d). The bubble size on each Cu 
site represents the conductance magnitude.    
The parameters $g_1=1.5t$ and $\mu=-0.4t$. 
}
\label{FIG:DISTRIBUTION}
\end{figure}

\begin{figure}
\caption{
The spatially-averaged differential tunneling conductance as a function 
of bias in a $d$-wave superconductor with each supercell containing   
one single impurity [i.e., the impurity concentration $c$ is about 
$0.08\%$] (solid line), a random distribution of impurities with 
$c = 1.28\%$ (dashed line), and $62.5\%$ of the randomly-distributed
 impurities of the former case changed to diagonal dimers 
(short-dashed line). The parameters $g_1=2t$ and $\mu=-t$. 
} 
\label{FIG:LDOS-AV}
\end{figure}


\begin{references}
\bibitem{ZBCPexpt} The most extensively studied type of experiments
 involves the observation of a zero-bias conductance peak in single
 particle tunneling. This type of experiments is reviewed in detail 
later in this work.
\end{references}
\end{document}